\documentclass[useAMS,usenatbib,12pt,epsfig,letter]{mn2e}
\usepackage[english]{babel}
\usepackage{amssymb}
\usepackage{graphicx}

\title
[Probing star formation with absorption line systems]
{Probing star formation across cosmic time with absorption line systems}
\author[B. M\'enard et al.]
{
\parbox[h]{\textwidth}{
Brice M\'enard$^{1}$, 
Vivienne Wild$^{2}$, 
Daniel Nestor$^{3,4}$,
Anna Quider$^3$, 
Stefano Zibetti$^5$} 
\vspace*{2pt} \\
\hspace{-.1cm}$^1$ Canadian Institute for Theoretical Astrophysics\\
\hspace{-.1cm}$^2$ Institut d'Astrophysique de Paris, C.N.R.S.\\
\hspace{-.1cm}$^3$ Institute of Astronomy, University of Cambridge\\ 
\hspace{-.1cm}$^4$ University of California in Los Angeles\\
\hspace{-.1cm}$^5$ Max Planck Institut f\"ur Astronomie\\
}

\def\mgii{Mg\,{\sc ii}}
\def\MgII{Mg\,{\sc ii}}
\def\HI{H\,{\sc i}}
\def \oii {[O\,{\sc ii}]}
\def \OII {[O\,{\sc ii}]}

\def \oiib {[O\,{\sc ii}]}
\def\d{\mathrm{d}}

\def\RAA{\rm \AA}
\def\sloii{\Sigma_{\rm L_{OII}}}

\newcommand{\begit}{\begin{itemize}}                                                                                               
\newcommand{\enit}{\end{itemize}}                                                                                                  
\newcommand{\begen}{\begin{enumerate}}                                                                                             
\newcommand{\enen}{\end{enumerate}}

\newcommand       \be           {\begin{equation}}                                                                                 
\newcommand       \ee           {\end{equation}}                                                                                   
\newcommand       \bea          {\begin{eqnarray}}                                                                                 
\newcommand       \eea          {\end{eqnarray}}

\newcommand{\beqa}{\begin{eqnarray}}                                                                                               
\newcommand{\eeqa}{\end{eqnarray}}

\def\la{\langle}                                                                                                                   
            
\begin{document}

\date{Draft, \today}

\maketitle

\begin{abstract}
We present an empirical connection between cold gas in galactic halos and star formation. Using a sample of more than 8,500 \MgII\ absorbers from SDSS quasar spectra, we report the detection of a $15\,\sigma$ correlation between the rest equivalent width $W_0$ of \MgII\ absorbers and the associated \oii\ luminosity, an estimator of star formation rate. 

This correlation has interesting implications: using only observable quantities we show that \MgII\ absorbers trace a substantial fraction of the global \oii\ luminosity density and recover the overall star formation history of the Universe derived from classical emission estimators up to $z\sim2$. We then show that the distribution function of \MgII\ rest equivalent widths, $\d N/\d W_0$  inherits both its shape and amplitude from the \oii\ luminosity function $\Phi(L)$. These distributions can be naturally connected, without any free parameter. 

Our results imply a high covering factor of cold gas around star forming galaxies: $C\gtrsim0.5$, favoring outflows as the mechanism responsible for \MgII\ absorption. We then argue that intervening \MgII\ absorbers and blue-shifted \MgII\ absorption seen in the spectra of star forming galaxies are essentially the same systems, implying that the observed outflowing gas can reach radii of $\sim50$ kpc. These results not only shed light on the nature of \MgII\ absorbers but also provide us with a new probe of star formation, in absorption, i.e. in a way which does not suffer from dust extinction and with a redshift-independent sensitivity. As shown in this analysis, such a tool can be applied in a noise-dominated regime, i.e. using a dataset for which emission lines are not detected in individual objects. This is of particular interest for high redshift studies.
\end{abstract}
\vspace{1cm}


\section{Introduction}

In the past twenty years, we have seen enormous advances in our understanding of galaxy evolution. Gravitational lensing has allowed us to interpret the properties of the host dark matter halos, population synthesis models can characterize the emission properties of galaxies over 3 orders of magnitude in stellar mass, and the star formation history of the Universe has been explored up to high redshift, revealing a peak at $z\sim2$. Along with this, the importance of feedback processes has emerged and one of the central questions which now needs to be addressed is how galaxies accrete, process and return gas into the inter-galactic medium (IGM).

Absorption line spectroscopy provides us with a powerful tool to address this question, to explore in detail the distribution of gas around galaxies and to study its interplay with star formation. It gives us access to a wide array of elements and allows us to detect low-density gas that is orders of magnitude below the detection threshold of most other techniques.

Historically, the study of QSO absorber-galaxy relationships began with \MgII\ absorption systems. This choice was mainly driven by practical constraints: among the dominant ions in \HI\ gas only \MgII\ could be studied from the ground at low redshifts ($z<2$). Following the first confirmation of a \MgII\ absorber-galaxy connection by \cite{1986A&A...155L...8B}, a considerable amount of work has focused on revealing the nature and origin of these systems. \MgII\ is known to be a tracer of galactic halos, routinely seen up to about 50\,kpc around galaxies. However, despite significant observational efforts in the last two decades, fundamental questions regarding the physical nature of the absorbing gas remain unanswered. Various models have been proposed in the literature: outflows \citep{2001ApJ...562..641B}, infalling gas \citep{2008ApJ...679.1218T}, extended disks \citep{2002ApJ...570..526S}, etc. but no consensus has been reached. Observationally, no compelling correlations between absorber properties and those of their associated galaxies have been reported. In particular, the physical process(es) governing the observed equivalent widths $W_0$, the basic observable quantity, and the origin of its distribution function are not yet understood. 

In this paper we present the detection of a correlation between the observed rest equivalent width of \MgII\ absorbers and the associated \oii\ luminosity and show that this relation has important implications. First, \mgii\ gas appears to trace a substantial fraction of the \oii\ luminosity density of the Universe. These absorber systems can therefore be used as a new probe for star formation over cosmological times. We show that the observed redshift dependence reproduces the decline in cosmic star formation rate since $z\sim2$ observed with emission line surveys. Finally, we derive a simple relation between the probability distribution function (p.d.f.) of $W_0$ and $L_{OII}$ and show that these two quantities are related.

This study was motivated by a number of recent results, indicating a possible link between the presence of strong \MgII\ absorbers and the star formation rate of their associated galaxies: \cite{Zibetti+07} analyzed the colors of stacked images of quasars with strong \MgII\ absorbers and reported that the best-fit spectral energy distributions (SED) require stronger emission lines for stronger \MgII\ absorbers. \cite{Wild07} created composite spectra of a subset of \MgII-selected absorbers and was able to detect their mean \oii\ emission. \cite{2009arXiv0907.0231R} studied the properties of a \MgII\ absorber-galaxy association and showed that the absorbing gas found at about 15\,kpc from the galaxy is consistent with having been ejected during the last burst of star formation of this galaxy. Finally, Nestor et al. (2010, in prep) studied the properties of two ultra-strong MgII absorbers ($W_0=5.6$ and $3.6$\RAA) and showed that the associated galaxies are in the starburst and post-starburst phase, respectively.

Using a different approach, by studying self-absorption in galaxy spectra, \cite{2009ApJ...692..187W} showed that all types of star forming galaxies trigger outflows seen through blue-shifted \MgII\ absorption. Such an analysis does not provide any information on the distance between the absorbing gas and the galaxy, and the connection to intervening absorber systems is not straightforward. However, we will show that our new results suggest a strong link between these two tracers of gas in and around galaxies. 

The outline of this paper is as follows: after briefly reviewing useful properties of \MgII\ absorbers and \oiib\ emission as a tracer of star formation, we describe the spectroscopic data analysis in Section~\ref{section_data}. We present the correlation between \oii\ luminosity surface density and \MgII\ equivalent width in Section~\ref{sec:results}. Section~\ref{sec:origin} presents the connection between $\Phi(L_{\rm OII})$ and the distribution of \MgII\ rest equivalent widths. Finally, we discuss the implications of these results in Section6 and summarize our work in Section 7. 

Throughout this paper we use the cosmology: $\Omega_M=0.3$, $\Omega_\Lambda=0.7$ and $H_0=70\,$km/s/Mpc.

\begin{figure*}
  \includegraphics[width=.44\hsize]{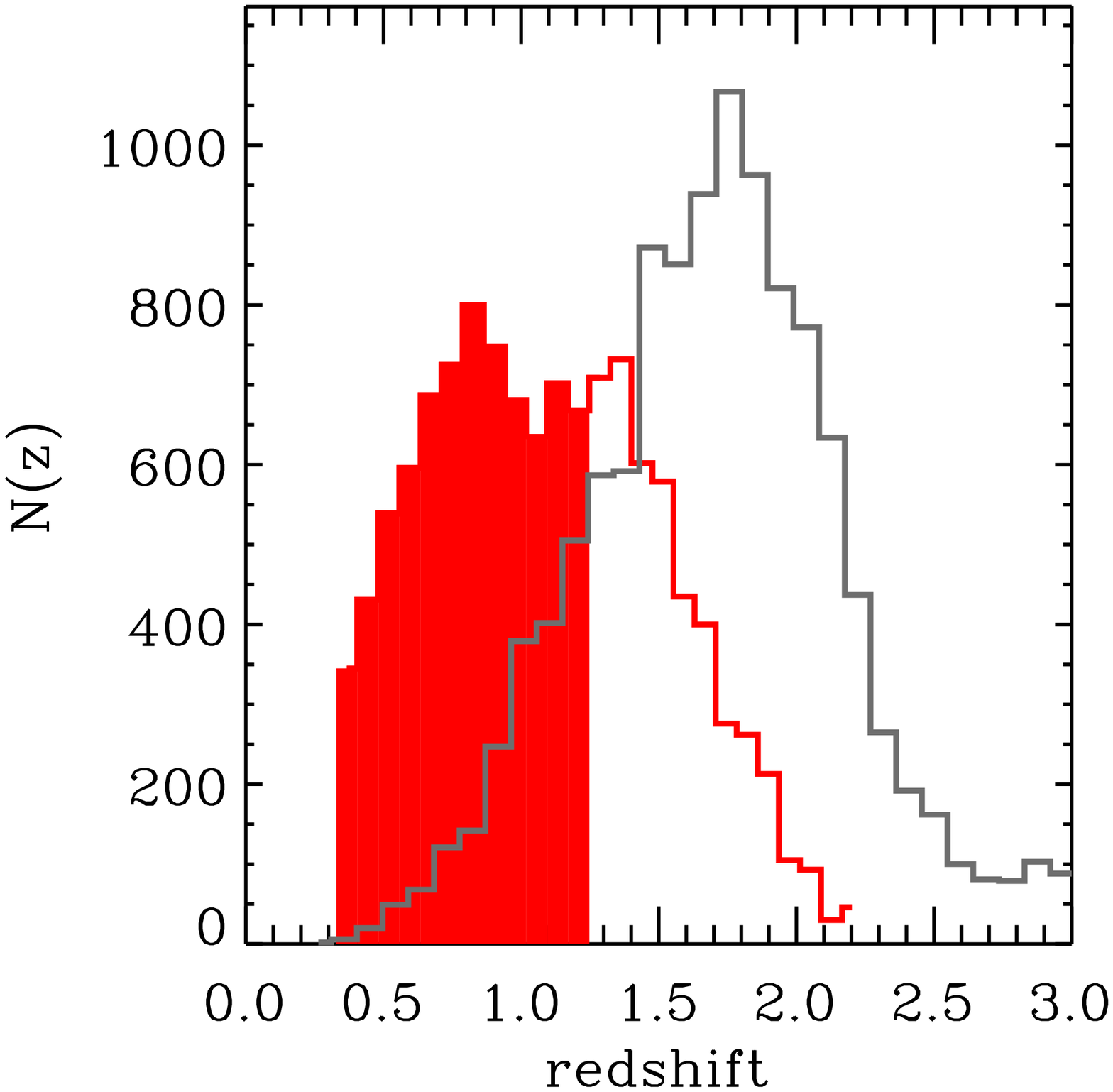}
  \includegraphics[width=.44\hsize]{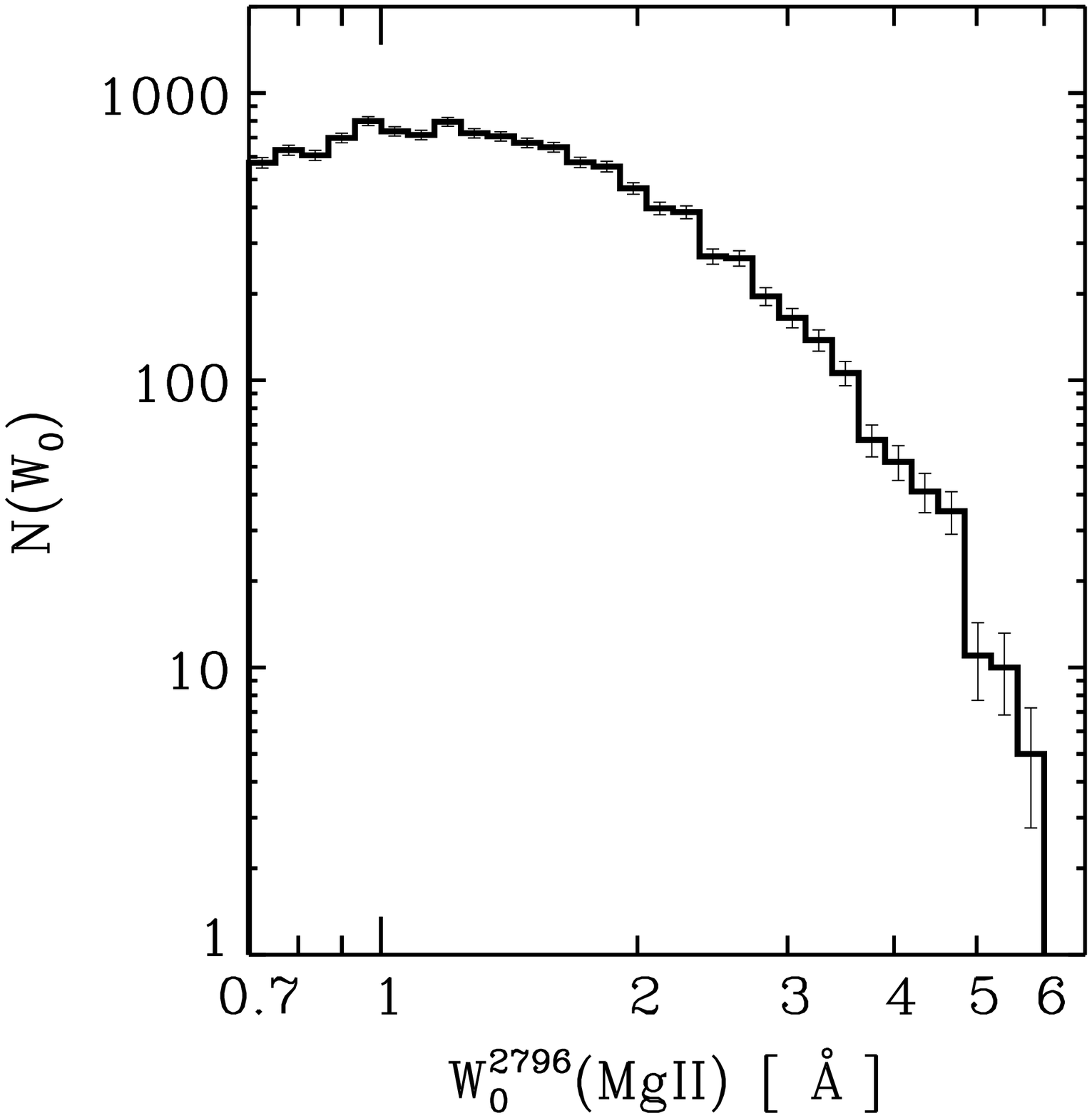}
\caption{\emph{Left:} redshift distributions of detected \MgII\ absorbers (red) and background quasars with $z<3$ (gray). The filled histogram represents the redshift range for which both \MgII\ and \oiib\ lines can be detected. \emph{Right:} observed distribution of MgII-$\lambda2796\,$\AA\ rest equivalent widths. Below 1 \AA\ the completeness becomes significantly lower than unity and ${\rm N}(W_0)$ departs from an exponential distribution.}
\label{plot_data}
\end{figure*}

\section{Astrophysical preliminaries}
\label{section:prelim}

\subsection{Properties of \MgII\ absorbers}
\label{sec:mgii}

Strong \MgII\ absorbers, classically defined with $W_0^{\lambda2796}>0.3\,{\rm \AA}$, are known to be associated with a range of galaxies, with $0.1\lesssim L \lesssim 10\,L^\star$ \citep{1991A&A...243..344B,1994ApJ...437L..75S,2007ApJ...658..185N} and are found at impact parameters ranging from a few to more than 50\,kpc \citep{Zibetti+07,Steidel+97,2008ApJ...679.1218T}. In the range $W_0^{\lambda2796} \gtrsim 1$\AA\ \MgII\ absorption lines are in general saturated \citep{Nestor+05} and no column density information can be directly extracted from the observed equivalent width, which can instead be used as an estimate of the overall gas velocity dispersion. High-resolution spectroscopic observations reveal that the absorption often originates from several velocity components \citep{2005pgqa.conf...24C}. Empirically, it has been found that in this regime the velocity dispersion of the gas follows $\Delta v \simeq 120\,(W_0/\RAA)\,$km/s (see \citealt{2006MNRAS.368..335E}, Fig. 3). 

A fundamental quantity describing the statistical properties of these absorbers is the distribution of \MgII\ rest equivalent width per unit redshift, ${\d^2 N}/{\d W_0\,\d z}$. This quantity has been accurately characterized using the Sloan Digital Sky Survey (SDSS) by \cite{Nestor+05}. These authors found that it is well described by an exponential distribution:
\begin{equation}
\frac{\d N}{\d W_0}= \frac{N^\star}{W^\star} ~  {\rm e}^{-W_0/ W^\star } 
\label{eq:Nestor}
\end{equation}
with the maximum likelihood values 
$N^\star=1.00\pm0.13\,(1+z)^{0.22\pm0.17}	$ and $W^\star=0.44\pm0.03\,(1+z)^{0.63\pm0.10}$. This result is valid down to a rest equivalent width limit of $\sim 0.3\,$\AA, below which the distribution follows a power-law behavior, making the overall distribution well described by a Schechter function. So far, no compelling model has been able to explain the origin of this distribution.

\subsection{\oii\ emission as a tracer of star formation}

Over the past 20 years many techniques have been explored to estimate the star formation rates (SFRs) of galaxies and, all together, have established the steep rise in the SFR density of the Universe from the present epoch to $z\sim2$ (\citealt{2006ApJ...651..142H} and references therein). One of the best-understood SFR indicators is H$\alpha\,\lambda6563$, whose luminosity is directly proportional to the hydrogen-ionizing radiation from massive stars (M$\gtrsim10\,M_\odot$) and therefore provides a near-instantaneous  ($\tau\lesssim10\,{\rm Myr}$) measure of the SFR with minimal dependence on the physical conditions of the ionized gas \citep{1998ApJ...498..541K}. Such observations are however limited to $z<0.4$ in the optical.
At higher redshifts, the \oiib\ $\lambda 3727$ nebular emission line has been used as an alternative SFR indicator (e.g. \citealt{1989AJ.....97..700G,1992ApJ...388..310K,2003ApJ...599..971H,2004AJ....127.2002K, 2005MNRAS.362.1143M}).

The \oii\ emission doublet is the shortest-wavelength strong emission line from low-density, photoionized galactic nebulae. The nebular lines effectively re-emit the integrated stellar luminosity short-ward of the Lyman limit. Only stars with masses $M>10\,{\rm M_\odot}$ and lifetimes $\tau<20\,{\rm Myr}$ contribute significantly to the integrated ionizing flux, so the emission lines provide a nearly instantaneous measure of the SFR, independent of the previous star formation history.
\oiib\ luminosity depends explicitly on the chemical abundance and excitation state of the ionized gas, and suffers a larger amount of dust extinction than H$\alpha$. Therefore, unlike the Balmer recombination lines, \oiib\ is not directly proportional to the SFR and must be calibrated either empirically \citep{1992ApJ...388..310K,2004AJ....127.2002K} or theoretically \citep{1997A&A...324..490B,
2001MNRAS.323..887C}. On average, it has been found that
\begin{equation}
\label{eq_kewley}
{\rm SFR~~ [ M_\odot/yr]} \simeq A  \times 10^{-41}~~ {\rm L_{OII}}\, [{\rm erg/s}]\,.
\end{equation}
The value of $A$ is of order unity, with the precise value depending on the broad band luminosity and dust content of the galaxy, etc. \citep{2006ApJ...642..775M}.

\section{DATA AND ANALYSIS}
\label{section_data}

Strong \MgII\ absorbers can be detected in SDSS quasar spectra over the redshift range $0.4\lesssim z\lesssim 2.2$. The spectroscopic observations make use of 3\arcsec\ fibers receiving photons originating from the entire path to the quasars (convolved with the seeing of the observations). The detection of a strong \MgII\ absorber indicates the presence of a galaxy close to the line-of-sight. If this galaxy is starforming, its narrow emission lines may be identified on top of the quasar spectral energy distribution. While the direct detection of these lines is possible only for a small number of absorber systems at low redshift (see \citealt{2009arXiv0912.0736N}), a statistical approach enables the measurement of the mean luminosity of  such lines well below the noise level of individual spectra and allow us to study the relationship between emission and absorption lines.

\subsection{The dataset}

The analysis makes use of a sample of \MgII\ systems compiled by \cite{Quider} using the method presented in \cite{Nestor+05} and extended to the SDSS DR4 dataset \citep{2006ApJS..162...38A}.  In this section we briefly summarize the main steps involved in the absorption line detection procedure. We refer the reader to \cite{Nestor+05} for more detail.

For each quasar spectrum in the SDSS DR4 database, \cite{Quider} fit a pseudo-continua in an iterative fashion using a combination of cubic-splines and Gaussians for both emission and absorption features.  The continuum-normalized SDSS QSO spectra were searched for \MgII$\lambda\lambda2796,2803$ doublets using an optimal extraction method employing a Gaussian line-profile to measure each rest equivalent width $W_0$. All candidates are interactively checked for false detections, a satisfactory continuum fit, blends with absorption lines from other systems, and special cases.  The identification of Mg II doublets required the detection of the 2796 line and at least one additional line, the most convenient being the 2803 doublet partner. A 5$\sigma$ significance level was required for all $\lambda2796$ lines, as well as a $3\sigma$ significance level for the corresponding $\lambda2803$ line.  Only systems 0.1c blue-ward of the quasar redshift and red-ward of Ly$-\alpha$ emission were accepted. From this catalogue we select systems where both \MgII\ and \oii$\lambda\lambda$3727,3730 are simultaneously covered by the accessible wavelength range. The final sample contains 8523 absorption line systems with $0.7<W_0<6.0\,$\AA\ and $0.4<z<1.3$.

The redshift distribution of the systems used in this analysis is presented in the left panel of Figure \ref{plot_data} as the filled histogram.  The right panel shows the \emph{observed} distribution of absorber rest equivalent width $W_0$.  As described above, the intrinsic distribution of rest equivalent widths follows an exponential distribution down to $W_0=0.3\,$\AA\ (see Eq.~\ref{eq:Nestor}).

\subsection{Measuring \oiib\ luminosities}
\label{section_composite}
\begin{figure}
  \includegraphics[width=\hsize]{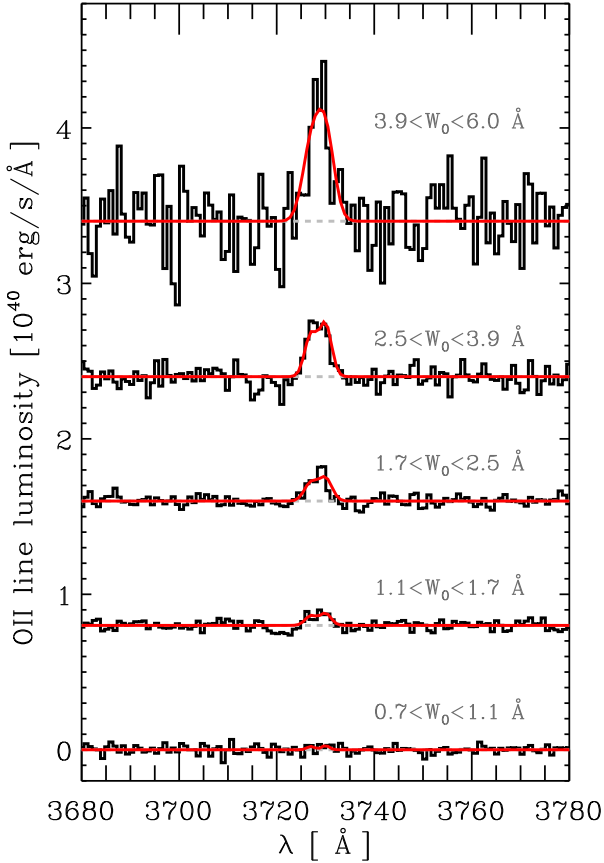}
\caption{Median flux in erg/s/cm$^2$/\AA\ from composite spectra in the region of the \oii\ emission line, for five bins of \MgII\ rest equivalent width. The red line is the best fit emission doublet.  The spectra have been offset for display purposes.}
\label{plot_OII}
\end{figure}

\subsubsection{Composite spectra and line measurement}

In general, the \oiib\ emission lines from galaxies with $z\gtrsim0.4$ are not detected in individual SDSS spectra. A small fraction of direct detections can be made at low redshift and will be discussed in future work. In the present analysis we estimate the average \oii\ emission of samples of \MgII\ absorbers by creating composite spectra in the absorber rest frame. In this procedure masks are applied to absorption lines expected at the redshift of all known \MgII\ absorber systems as well as sky emission lines. The continuum is defined with an iterative running median of sizes ranging from 500 to 15 pixels. Narrow features caused by unidentified systems with $z\ne z_{abs}$ cannot be masked with a reliable uniformity, and are thus treated as an additional noise contribution. Each continuum-subtracted spectrum is converted into units of luminosity surface density (see below) and samples of spectra are combined to create average and median composites. In general the results from these two estimators are found to be similar. Since the former ones are usually noisier we will focus only on median values below.

In each composite spectrum the \oiib$\lambda\lambda$3727,3730 doublet was fit with a double Gaussian with line ratio fixed at 4:5 and freely varying line width. Line luminosity surface densities (see below) were measured from the fitted Gaussian parameters.  In order to estimate the noise level of the line detection we applied the same fitting procedure to the region of the spectrum $3500<\lambda_{rest}<3600\,\rm{\AA}$ where no strong emission/absorption line is expected. 

Examples of average composite spectra around the \oii\ line are shown in Fig. \ref{plot_OII} for samples of \MgII\ absorbers with different rest equivalent widths. We note that no shift in the centroid of the \oii\ lines with respect to those of the \MgII\ lines is detected. These results indicate a strong correlation between \oii\ emission and \MgII\ absorption rest equivalent width. This correlation will be analyzed quantitatively below.

\subsubsection{Aperture correction}\label{sec_aper}

The available sample of absorbers for which both \MgII\ and \oii\ lines are simultaneously accessible spans a substantial redshift interval: from 0.4 to 1.3. In this range, the physical area corresponding to a given angular aperture varies by a factor of about 2.5. This is illustrated in Fig. \ref{plot_loii_fiber} in the Appendix in which we show the physical size covered by the 3$\arcsec$ diameter SDSS fibre aperture as a function of redshift. 
In order to combine coherently, or compare, signals from different redshifts, we estimate the \oii\ luminosity surface density, $\sloii$. For each \MgII\ absorber, we thus convert the spectrum from flux units into luminosity units, at the redshift of the absorber. We then divide by the surface area of the SDSS fibre at this redshift, to obtain luminosity surface density:
\be
\sloii(z) = \frac{L_{OII}(z)}{\Omega_{f}\, D_A^2}\,,
\ee
where $\Omega_{f} = \pi \theta_f^2$ and $\theta_f$ is the angular radius of the SDSS fibre and $D_A(z)$ is the angular diameter distance at the redshift of the absorber.

\begin{figure*}
  \includegraphics[width=.49\hsize]{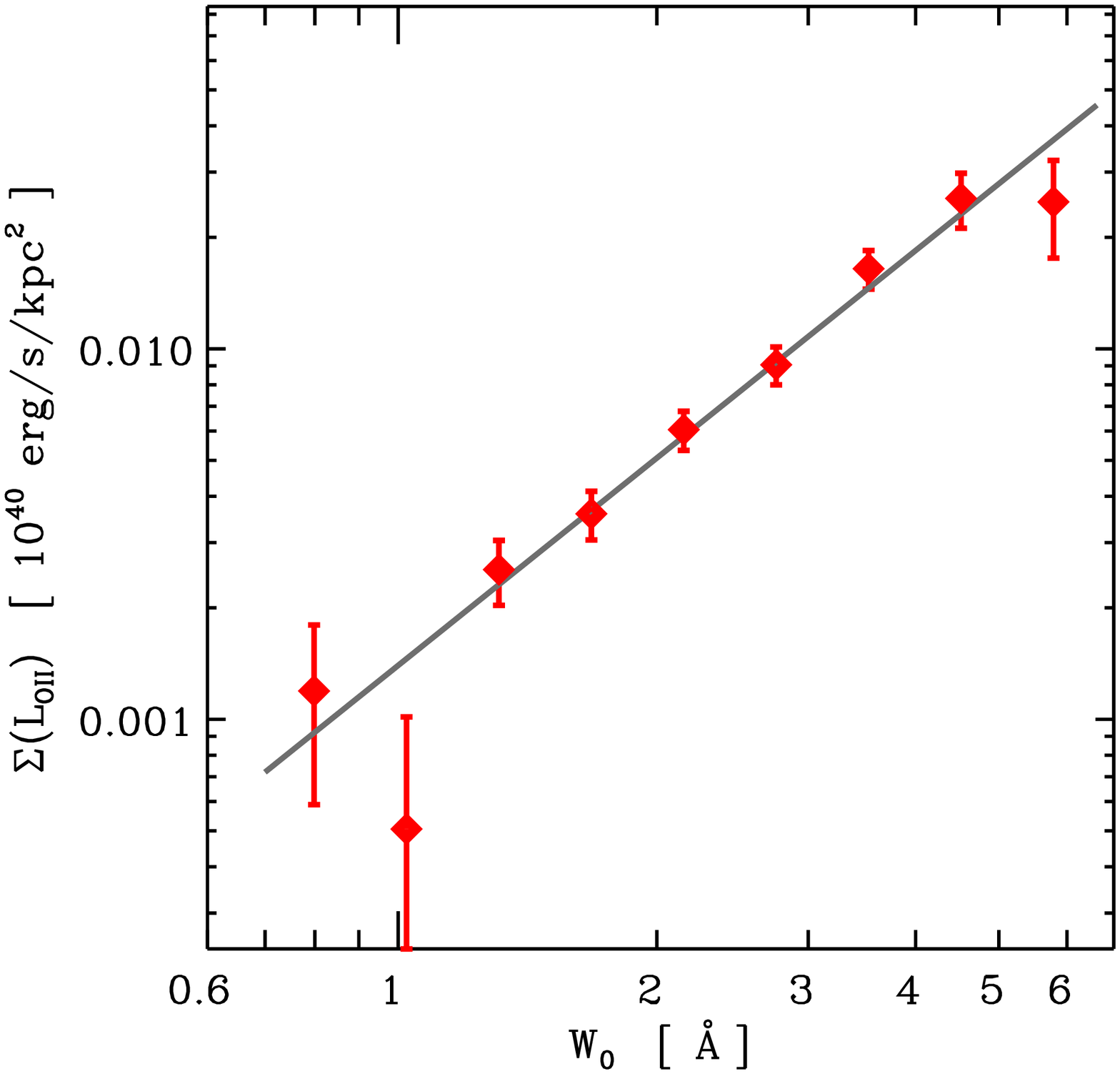}
  \includegraphics[width=.49\hsize]{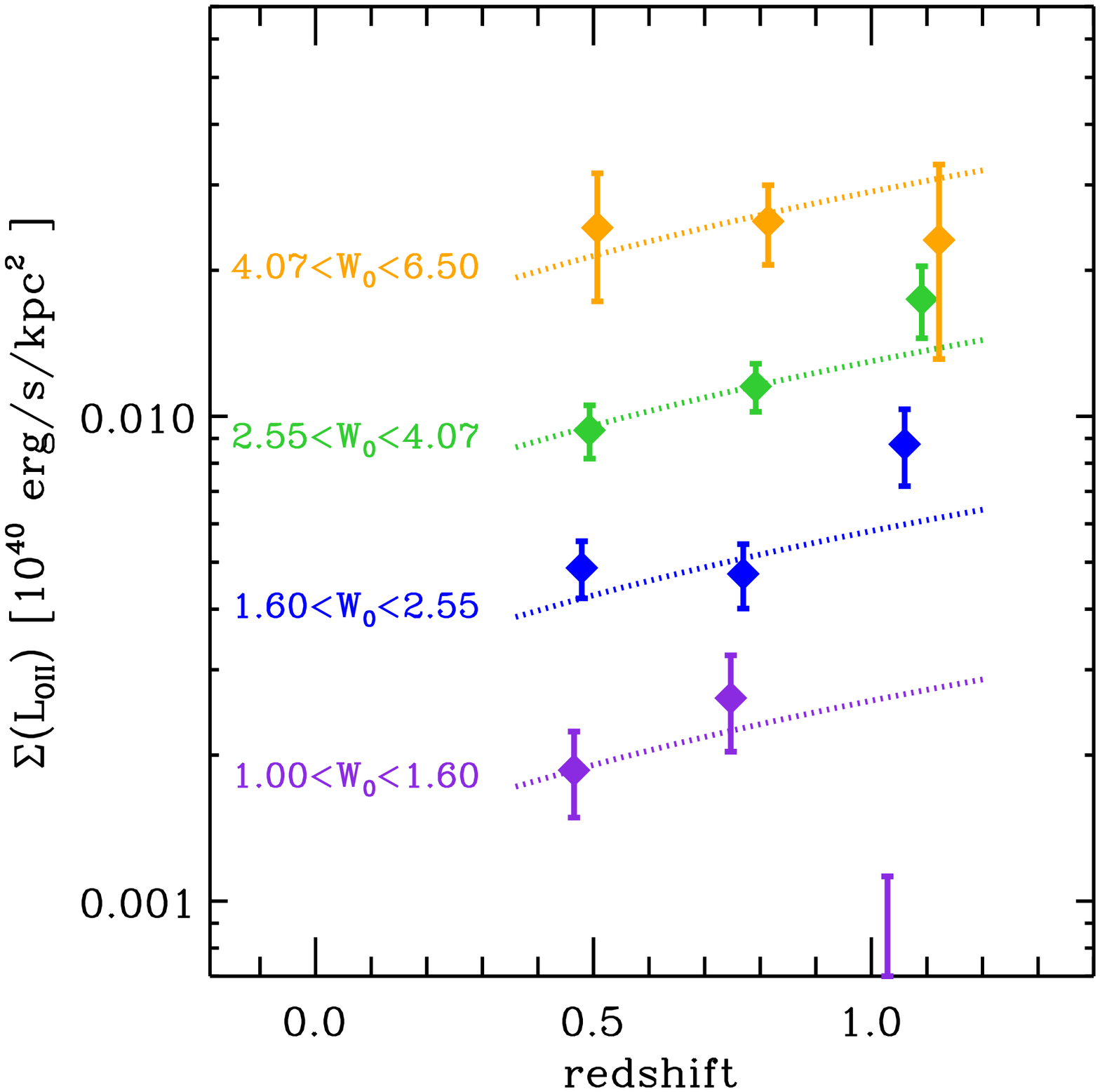}
  \caption{\emph{Left:} mean \oii\ luminosity surface density, $\sloii$ (averaged within the 3-arcsec fibre) as a function of \MgII\ rest equivalent width $W_0$. The solid line is the best fit power-law. \emph{Right:} mean \oii\ luminosity surface density measured in three redshift bins, as a function of $W_0$. The dashed lines show the best 3-parameter fit to the entire sample.
}
\label{plot_sigma_oii}
\end{figure*}

\section{\MgII\ absorption as a tracer of star formation}
\label{sec:results}

In this section we use \MgII\ absorbers as tracers and show they can be used as a probe of star formation. This is done independently of any relation between \MgII\ absorbers and galaxies.

\subsection{\MgII\ - \oiib\ correlation}
\label{sec:correlation}

Using a logarithmic binning in \MgII\ rest equivalent width, $W_0^{\rm MgII}$, we create composite spectra in the absorber rest-frame and estimate $\sloii$, the median \oii\ luminosity surface density. The observed relation between these two parameters is presented in Fig. \ref{plot_sigma_oii} with red data points. \MgII\ absorbers with equivalent widths $0.7<W_0<6\,{\rm\AA}$ are seen to strongly correlate with \oii\ emission, spanning more than a factor 30 in luminosity surface density. The observed relation, averaged over the redshift range $0.4<z<1.3$, is simply described by
\begin{equation} 
\label{eq_sigoii}
\langle\, \Sigma_{\rm L_{OII}}(W_0) \,\rangle = A \left( \frac{W_0}{1\,{\rm \AA}} \right)^{\alpha}
\end{equation}
with $A=(1.48 \pm 0.18) \times 10^{37}\,{\rm erg\,s^{-1}\,kpc^2}$ and $\alpha=1.75\pm0.11$. This fit is shown in Figure \ref{plot_sigma_oii} with the solid line. 

This result shows that, without any prior knowledge of the underlying galaxy distribution, \MgII\ absorbers can be used to trace the distribution of \oii\ emission, i.e. star formation.
This property is interesting since the detectability of absorber systems does not depend on redshift and is not significantly affected by dust extinction. They can therefore provide us with a powerful probe of star formation over cosmological times. This will be discussed in more detail below.

\subsubsection{Redshift dependence}

\begin{figure*}
  \includegraphics[width=.48\hsize]{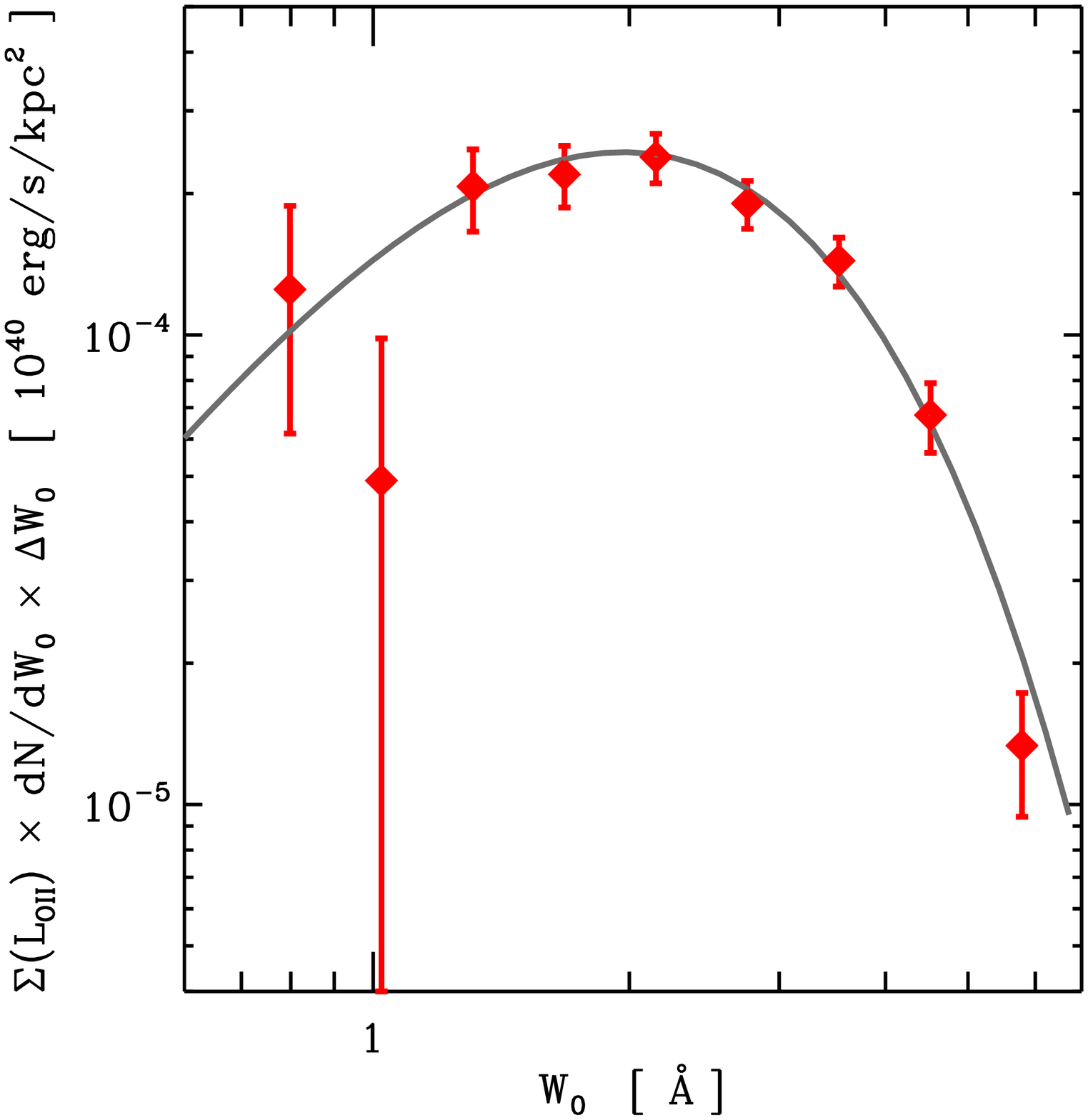}
    \includegraphics[width=.48\hsize]{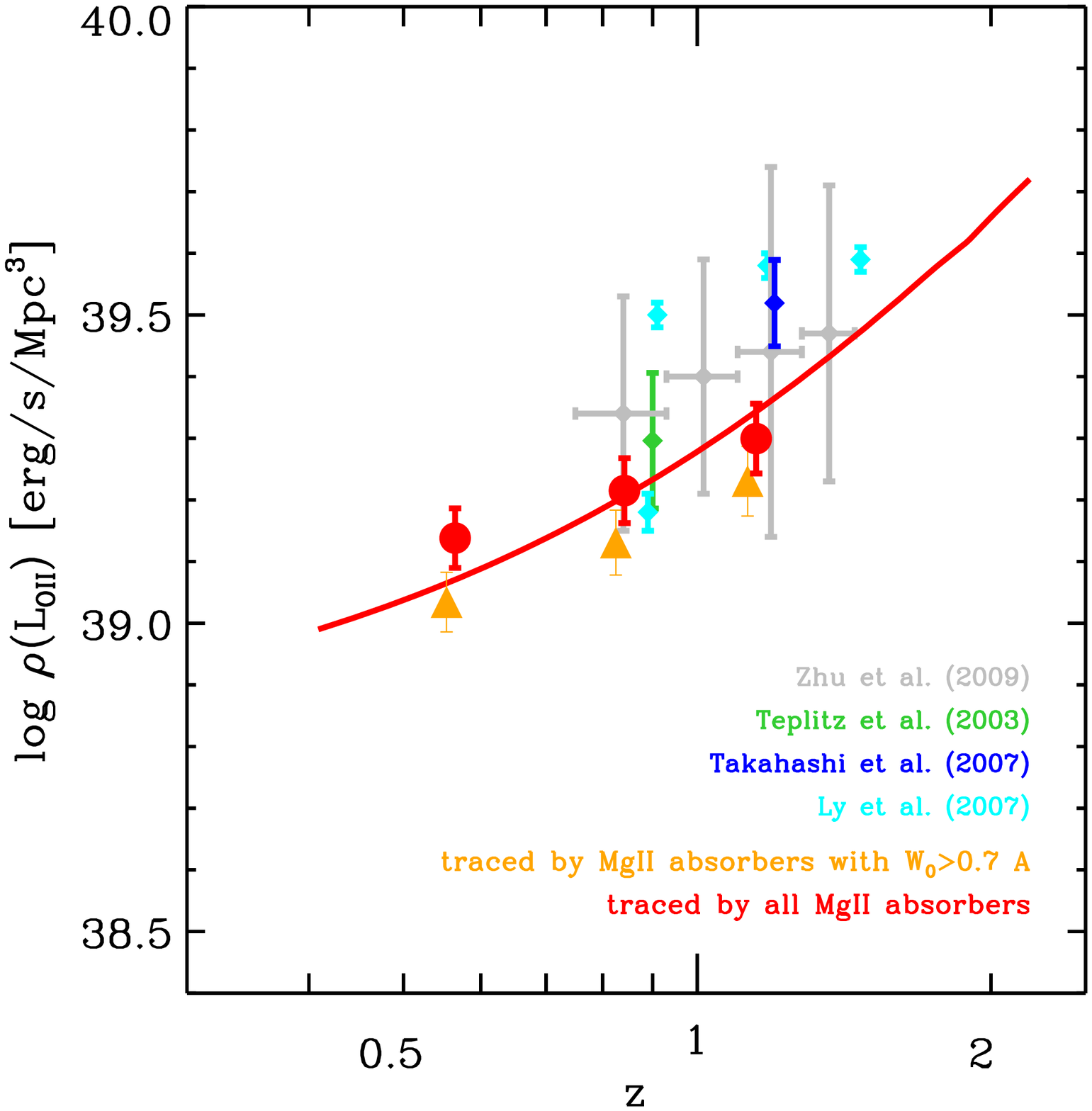}
  \caption{  \emph{Left:} Relative contribution of \oii\ luminosity surface density, $\sloii$ as a function of \MgII\ rest equivalent width $W_0$. The solid line is the best fit given by Eq.~\ref{eq:rho_oii}. \emph{Right:} the \oii\ luminosity volume density from \MgII\ absorbers (with $W_0>0.7\,\RAA$ in orange and all systems in red) computed at three different redshifts. The solid line shows the best fit trend computed within the range for which the incidence of \MgII\ absorbers ($\d N/\d z$) is measured. The other data points show direct \oii\ luminosity density measurements from the literature.
Note that no dust correction has been applied to any of these \OII\ luminosity measurements, allowing direct comparison between samples. \MgII\ absorbers appear to trace a substantial fraction of the global \oiib\ luminosity density in the Universe.}
\label{plot_oii_stats}
\end{figure*}

In order to investigate the redshift dependence of the above relation, we measure the mean \OII\ luminosity surface density for three redshift intervals: $0.36<z_1<0.67<z_2<0.99<z_3<1.3$. The results are shown in the right panel of Fig. \ref{plot_sigma_oii} where each color corresponds to a given range of \MgII\ rest equivalent widths, resulting in 12 independent data points. Remarkably, they appear to be distributed regularly in the $L_{\rm OII}-z$ plane. This property further demonstrates that the relation between \MgII\ absorption properties and observed \OII\ luminosity surface density holds over a significant cosmological time span.  This figure illustrates that \MgII\ absorbers 
are found, on average, in regions having a predictable value of \OII\ luminosity surface density. This correspondence is valid over a large range in both redshift and \OII\ luminosity.

A global fit to all the data points in the redshift-$W_0$ plane gives:
\begin{equation}
\label{eq:fit_global}
\langle \sloii(W_0,z) \rangle = {\rm A}\,\left(\frac{W_0}{1\,{\rm \AA}}\right)^\alpha\,(1+z)^\beta
\end{equation}
with ${\rm A}=(7.9\pm2.0)\times10^{36}\,{\rm erg\,s^{-1}\,kpc^2}$, $\alpha=1.7\pm0.1$ and $\beta=1.1\pm0.4$.
The results of this global fit are shown in the right panel of Fig. \ref{plot_sigma_oii} as dotted lines. Using this simple formula, we are able to fit 12 data points with 3 parameters resulting in a reduced $\chi^2$ of $1.3$.
The redshift dependence of this relation is remarkably weak. The mean \oiib\ luminosity surface density changes by less than 50\% over a unity redshift range, corresponding to a timescale of about 5 Gyr. Understanding the origin of this dependence would be of particular interest and would allow us to directly use \MgII\ absorbers as a probe of star formation without the need of detecting \oiib\ emission.

\subsection{OII luminosity density}

The SDSS fibers used to observe quasars can be considered as a set of cones integrating light emitted along the line-of-sight. Each quasar spectrum therefore probes a given volume and the set of (high-redshift) quasar spectra can be considered as a random sampling of the low-redshift Universe. Within this volume, we estimate the \oiib\ luminosity density, $\mathcal{L}_{OII}$, traced by \MgII\ absorbers. Note that we are only interested in estimating a density and therefore do not need to include any emission contribution originating outside the fibers. We then use $\mathcal{L}_{OII}$ to estimate the star formation rate density probed by these systems as a function of redshift and compare it to the total value in the Universe.

The co-moving \oii\ luminosity density probed by \MgII\ absorbers is given by
\bea
{\mathcal L_{OII}}(z) &=&\frac{L}{V_C}\nonumber\\
&=& \frac{1}{(1+z)^2}\,\frac{\d L}{D_A^2(z)\,\d \Omega}\;\frac{\d N}{\d r}
\label{eq:}
\eea
where $L$ is the \oii\ luminosity associated with absorbers observed within the co-moving volume $V_C$, which is accessible along the pathlength towards the background quasars (with and without absorbers). In the second expression, $D_A^2(z)\,(1+z)^2\,\d \Omega$ is the co-moving surface of the aperture, ${\d N}/{\d r}$ is the mean number of absorbers expected within a pathlength $\d r$ and $\d L$ is the \oiib\ luminosity observed within the corresponding volume element. Introducing the cosmology-dependent function
\be
{\mathcal E}(z)=\frac{\d z}{\d r}\,\frac{1}{(1+z)^2}= \frac{H_0}{c}\frac{\sqrt{\Omega_{\rm M} (1+z)^3 +   \Omega_\Lambda}} {(1+z)^2}\;
\label{eq:E}
\ee
we can simply express the luminosity density probed by a population of \MgII\ absorbers as a function of observable quantities
\bea
\mathcal{L}_{OII}(z) &=&\sloii(z)\;\frac{\d N}{\d z}\;{\mathcal E}(z)\,,
\label{eq:oii_density}
\eea
where $\d N/\d z$ is the line-of-sight incidence of absorber systems and $\sloii$ the mean luminosity surface density associated with them. Eq.~\ref{eq:oii_density} indicates that, for a given cosmology, \emph{the \OII\ luminosity density traced by \MgII\ absorbers can be derived without any assumption}. In particular it does not depend on the spatial extent of the gas and the relation between absorbers and galaxies. A similar formalism was used by \cite{2003ApJ...593..235W} and \cite{Wild07} to estimate the contribution of damped Lyman-$\alpha$ systems and CaII absorption systems to the overall star formation rate.

In order to account for the varying completeness of the absorber sample with $W_0$, we estimate Eq.~\ref{eq:oii_density} by integrating over the intrinsic rest equivalent width distribution
\be
\mathcal{L}_{OII}(z) = {\mathcal E}(z)\;\int \d W_0\;\frac{\d^2 N}{\d W_0\,\d z}\;\sloii(W_0,z)\;
\label{eq:rho_oii}
\ee
where ${\d^2 N}/{\d W_0\,\d z}$ is the completeness-corrected distribution of rest equivalent widths given by \cite{Nestor+05}, see Eq.~\ref{eq:Nestor}. The relative contribution of the integrand is shown in the left panel of Fig.~\ref{plot_oii_stats} for the sample spanning the entire redshift range. It shows that most of the \OII\ luminosity surface density, and therefore most of the star formation, is associated with systems having $W_0\sim2\,{\rm \AA}$. Interestingly, the same result was obtained regarding the dust content of \MgII\ absorbers \citep{2008MNRAS.385.1053M}. We will return to this point in the discussion. We note that estimating the integral in Eq.~\ref{eq:rho_oii} from the lowest rest equivalent width available in this analysis, i.e. $W_0=0.7\,\RAA$, gives a result only 15\% lower than integrating from zero with the extrapolated fitting functions at the mean redshift of the sample.

We compute $\mathcal{L}_{OII}$ using Eq.~\ref{eq:rho_oii} for the three redshift bins defined above and the results are given in Table~\ref{table:rho_oii}. The results are presented in the right panel of Figure~\ref{plot_oii_stats}. The orange data points show direct measurements for our sample of \MgII\ absorbers with $W_0>0.7\,\RAA$ and the red ones show the estimate for all \MgII\ absorbers, i.e. including the contribution from weaker systems. For comparison, we show a compilation of measurements of the total \oiib\ luminosity density obtained from narrow band filter and emission line surveys. Note that no dust correction has been applied to any of these datasets, allowing direct comparisons between observed quantities. This figure shows that, at $z\sim 1$, \emph{the amount of \OII\ luminosity density traced by \MgII\ absorbers is a substantial fraction of the total density in the Universe} estimated from direct emission measurements. 
Given the non-negligible scatter in the reported values of $\mathcal{L}_{OII}$ from photometric and spectroscopic surveys at around $z=1$, it is not yet possible to precisely quantify the fraction of \oii\ luminosity traced by \MgII\ absorbers. Nevertheless our results show that at least half of the \oii\ luminosity density is traced by strong \MgII\ absorbers. This implies that a high fraction of 
star-forming galaxies are associated with strong \MgII\ absorbers and vice-versa. In other words, the presence of \MgII\ absorbers around a galaxy depends on its (recent) star formation rate.

To further illustrate this connection, we explore the redshift dependence of $\mathcal{L}_{OII}$. As shown above, this quantity is given by the product of two observables: 
${\d^2 N}/{\d W_0\,\d z}$, which has been accurately measured up to $z=2.2$ (see Eq.~\ref{eq:Nestor}) and $\sloii(W_0,z)$ which is parametrized in Eq.~\ref{eq:fit_global} up to $z=1.3$.
In order to estimate $\mathcal{L}_{OII}$ up to $z=2.2$, we use an extrapolation of Eq.~\ref{eq:fit_global}, a relation which is weakly redshift dependent. The result is shown in Fig.~\ref{plot_oii_stats} with the solid line. The overall redshift dependence appears to be in agreement with the trend given by direct \oii\ emission surveys. 


\begin{table}
\begin{center}
\begin{tabular}{ccc}
\hline
\hline
redshift &   $\mathcal{L}_{OII}$ for $W_0>0.7\,\RAA$ &   $\mathcal{L}_{OII}$ for all $W_0$ \\
interval &  [${\rm 10^{39}\, erg\,s^{-1}\,Mpc^{-3}]}$ &  [${\rm 10^{39}\, erg\,s^{-1}\,Mpc^{-3}]}$\\
\hline
   $0.36<z<0.67$ & $1.08\pm0.13$ & $1.37\pm0.16$ \\   
   $0.67<z<0.99$ & $1.35\pm0.17$ & $1.64\pm0.21$\\
   $0.99<z<1.30$ & $1.70\pm0.24$ & $2.00\pm0.28$\\
\hline \hline
 \end{tabular}                                                                                                                                
\caption{\OII\ luminosity density traced by \MgII\ absorbers as a function of redshift. No dust corrections have been applied.}
\label{table:rho_oii}
\end{center}
\end{table}

\subsection{Star formation history}

\begin{figure}
    \includegraphics[width=\hsize]{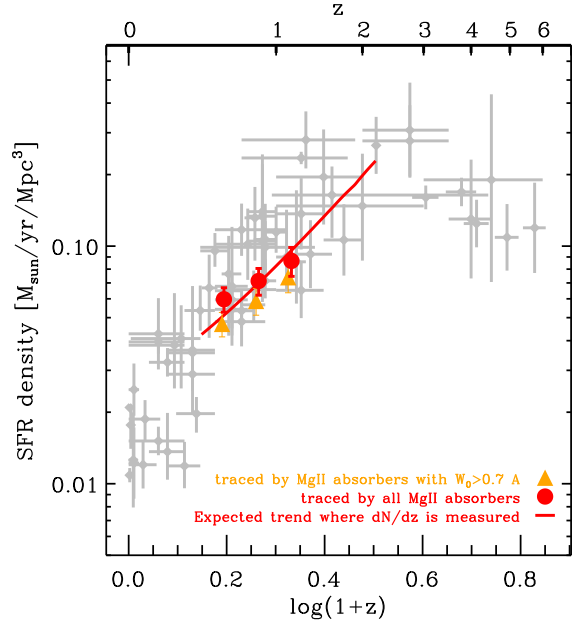}
  \caption{Star formation rate density as a function of redshift. The gray data points are from the Hopkins (2004) compilation and use various types of estimators. The orange and red data points show the SFR density traced by $W_0>0.7\,\RAA$ and all \MgII\ absorbers. The red curve shows the SFR density estimate using $\d N/\d z$ measured up to $z=2.2$ and the extrapolated \oii-\MgII\ relation.}
\label{plot_madau}
\end{figure}
We now show how the global star formation probed by \MgII\ systems compares to the overall star formation history in the Universe estimated by various techniques. As mentioned above, \oii\ luminosity can be used as an estimator of star formation rate, however the scaling between these two quantities is not straightforward: it is based on empirical relations, calibrated at various redshifts and subject to dust extinction corrections \citep{1998ApJ...498..541K, 2006ApJ...642..775M}. Here we choose to apply the average scaling coefficient used in a recent analysis by Zhu et al. (2009) for galaxies with $0.75<z<1.45$ (see their Table 2). These authors made use of the empirical correlation derived by Moustakas et al. (2006) between the absolute B-band magnitude, and the L\OII/SFR ratio. This calibration statistically accounts for the gross systematic effects of reddening, metallicity, and excitation, all of which correlate with optical luminosity, and has been shown to work reasonably well for star-forming galaxies at $0.7 < z < 1.4$ (Moustakas et al. 2006; Cooper et al.2008). Averaging their scaling coefficients over the redshift interval $0.75<z<1.45$, we get
\be
\dot \rho_\star = 4.2\;10^{-41}\times {L}_{OII} \;{\rm M_\odot}\,{\rm yr}^{-1}\,{\rm Mpc}^{-3}
\ee
where ${L}_{OII}$ is in erg s$^{-1}$. With the limitations of such a scaling in mind, we convert our \oii\ luminosity density estimate into a star formation rate density and present the results in Figure~\ref{plot_madau}. The gray data points show the estimates summarized in the Hopkins (2004) compilation\footnote{
The more recent compilation by Hopkins et al. (2006) focuses on star formation measurements derived from rest-frame UV observations rather than emission-line based star formation estimates. It is therefore less appropriate for our comparison.
}.
We show the estimated star formation rate density from \MgII\ absorbers with $W_0>0.7\,\RAA$ and all systems with the orange and red data points respectively. The solid line shows the best fit trend obtained in Eq.~\ref{eq:rho_oii}, using the \oii\ - \MgII\ scaling given in Eq.~\ref{eq:fit_global} and $\d N/\d z$ measurements available up to $z=2.2$. As can be seen, both the overall amplitude and the redshift dependence are in good agreement with other star formation rate estimators. The global star formation probed by \MgII\ absorbers can therefore be used as an independent probe of the star formation \emph{history} in the Universe. 

Interestingly, extrapolating the comparison between the SFR density traced by \MgII\ absorbers and observations of the global SFR density to higher redshifts suggests that the product $\sloii\times{\d N}/{\d z}$ (see Eq.~\ref{eq:oii_density}) should reach a maximum value at around redshift two and then decrease at higher $z$. Since $\sloii(z,W_0)$ appears to be weakly redshift dependent, the expected decrease in this product should be driven by a decrease in ${\d N}/{\d z}$ beyond redshift two. This can be tested observationally using near-infrared spectroscopy of an ensemble of quasars, with instruments such as X-shooter \citep{2006SPIE.6269E..98D} or the upcoming FIRE spectrograph \cite{2008SPIE.7014E..27S}.


\section{The \oii\ luminosity function and the distribution of \MgII\ rest equivalent widths}
\label{sec:origin}

\begin{figure*}
  \includegraphics[width=.49\hsize]{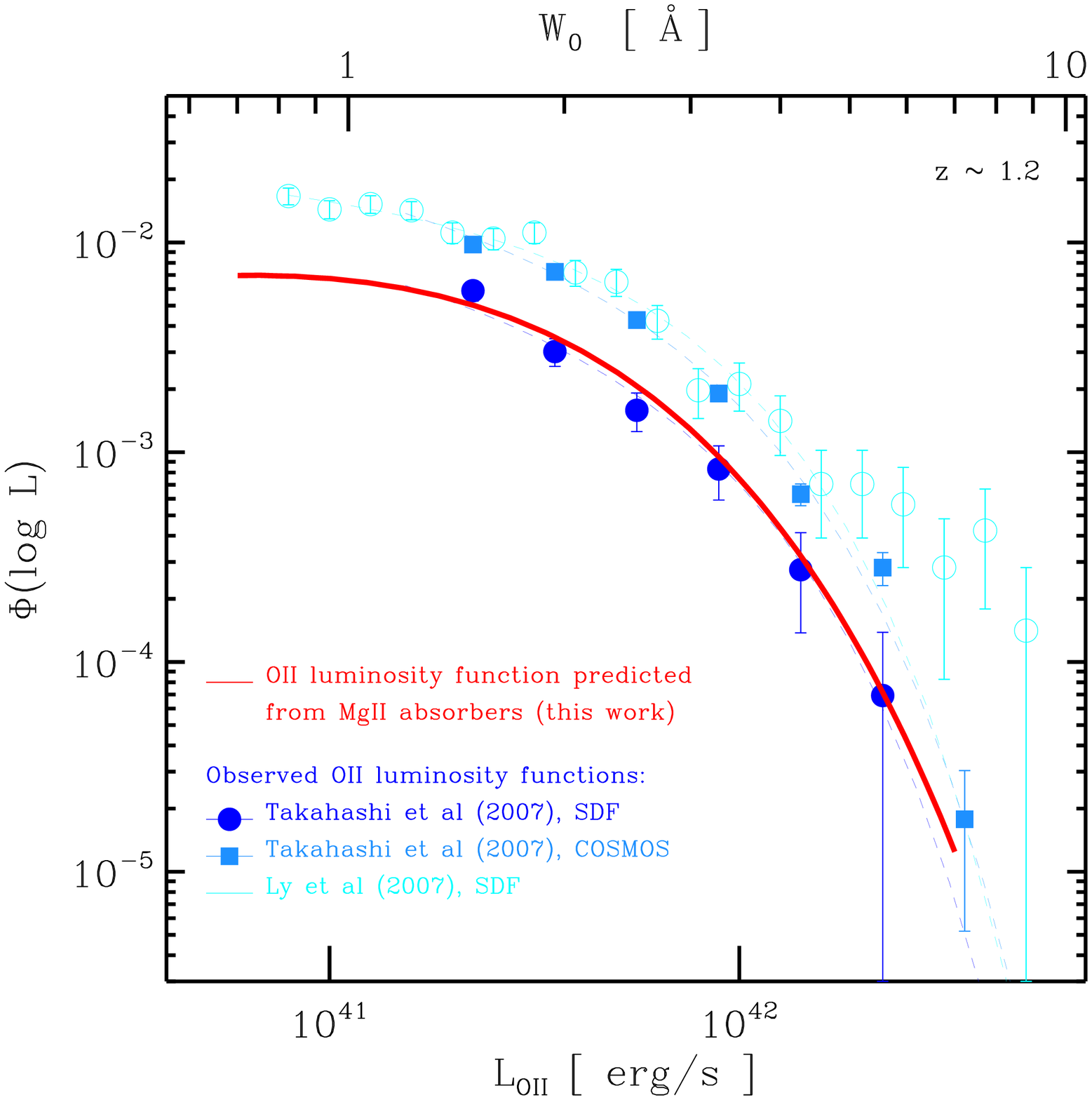}
    \includegraphics[width=.49\hsize]{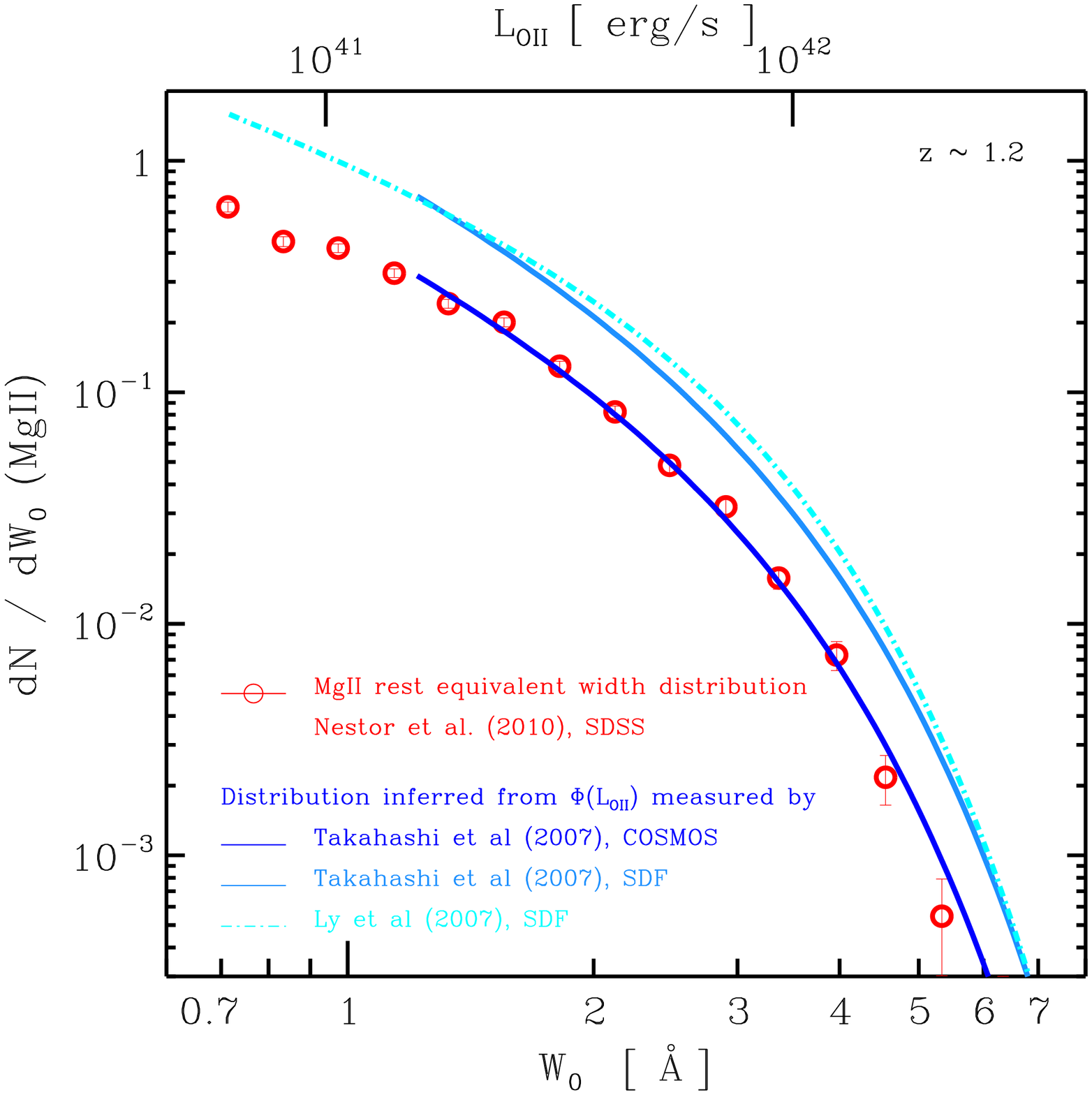}
  \caption{The connection between the \oii\ luminosity function, $\Phi(L)$, and the distribution of \MgII\ rest equivalent width, ${\d {\rm N}}/{\d W_0}$.
  \emph{Left:} $\Phi(L)$ derived from the distribution of \MgII\ absorbers (solid red) using the observed scaling relation between \oii\ emission and \MgII\ absorption, ${\d {\rm N}}/{\d W_0}$ estimated at $z\simeq 1.2$ from the SDSS (Nestor et al. 2005) and a cross-section with a radius of 50 kpc. Three observed \oii\ luminosity functions are shown for comparison. No dust corrections are applied allowing direct comparisons.
  \emph{Right:} \MgII\ rest equivalent width distribution at $z\simeq1.2$ measured with the SDSS (red data points, Nestor et al. 2010). The three blue-colored curves show our estimates of ${\d^2 {\rm N}}/{\d W_0\,\d z}$ based on the observed \oii\ luminosity functions.
}
\label{plot_oii_mgii_connection}
\end{figure*}

In the previous section we made use of \MgII\ absorbers as tracers and showed that, independently of their relations to galaxies, they can be used as a probe of star formation in the Universe. The connection between \MgII\ absorbers and star formation can be explored one step further. We now show that the distribution of \MgII\ rest equivalent widths ${\d N}/{\d W_0}$, and the \OII\ luminosity function, i.e. a measure of the probability distribution function (p.d.f.) of star formation rate in the Universe, are connected, both in amplitude and shape. Our goal here is only to reveal the existence of this relation and explore its implications. A more detailed comparison and study of possible deviations is left for future work. Moreover, as explained below, the scatter in the reported values of the \oii\ luminosity function (see Fig.~\ref{plot_madau} and \ref{plot_oii_mgii_connection}) currently limits the level to which we can meaningfully refer to the amplitude of the \oii\ luminosity function.

Given the existence of a relation between the mean \oii\ luminosity surface density and $W_0$ we can express, in a statistical sense, the \OII\ luminosity function $\Phi(L_{\rm OII})$ in terms of $W_0$(\MgII). We can write:
\bea
\Phi(\log L) &=& \frac{\d^2 N}{\d \log L\,\d V}\;\frac{1}{ C} \nonumber\\
&=& \frac{\d^2 N}{\d W_0\,\d z} \; \frac{\d W_0}{\d \log L}\;\frac{{\mathcal E}(z)}{C\,\sigma}
\label{eq:phi}
\eea
where $L$ represents the mean \OII\ luminosity per galaxy traced by a strong \MgII\ absorber, $\sigma$ is the cross-section for absorption and $C$ is the  covering factor for \MgII\ absorption within this cross-section, which can be expressed as the product of the fraction of galaxies giving rise to \MgII\ absorption and the spatial cross-section taking into account the angular variations and clumpiness of the gas: $C=C_f\,C_\Omega$. The function ${\mathcal E}(z)$ carries the dependence on cosmology, as defined in Eq.~\ref{eq:E}. 

In order to show the connection between the distribution of \MgII\ rest equivalent widths and the \oii\ luminosity function, we use a simplifying assumption: we consider that the cold gas distribution around star forming galaxies is isothermal, as motivated by \cite{2008ApJ...683...55C}. In this case, the 2-dimensional density distribution goes like $\rho_{\rm 2D}\propto b^{-1}$ and the probability of observing a system at an impact parameter $b$, $\rho_{\rm 2D}(b)\,b\,\d b$, is constant up to a maximum impact parameter $b_{\rm max}$. The uniform cross-section is then simply given by $\sigma=\pi\,b_{\rm max}^2$. The fraction of galaxy light falling within the fiber of projected radius $r_f$ is, on average, given by $(r_f/b_{\rm max})^2$ and the mean \oii\ luminosity per \MgII\ galaxy is $\rm{L_{OII,gal}}  = \sloii\;\sigma$, where $\sloii$ is the mean \oii\ luminosity surface density derived above. This allows us to express the luminosity function as:
\bea
\label{eq:phi_simplified}
\Phi(\log L) &=& \frac{\d^2 N}{\d W_0\,\d z}\; \frac{\d W_0}{\d \log (\sloii)} \;\frac{{\mathcal E}(z)}{C\,\sigma}\nonumber \\
&=& \frac{\d^2 N}{\d W_0\,\d z}\; \frac{{\mathcal E}(z)}{\alpha\;C\,\sigma}\,,
\eea
with $\alpha=\d \log \sloii/\d W_0$. Using this approximation, we now compute the \oii\ luminosity function expected from \MgII\ absorbers, as a function of $W_0$. The first two terms of the above equation are observable quantities, described by Eq.~\ref{eq:Nestor} and \ref{eq:fit_global}. Various observations of \MgII\ absorbers-galaxy associations indicate that the gas extends up to projected radii of about 50 kpc (Steidel et al. 1995, Zibetti et al. 2007, Chen \& Tinker 2008). We will therefore use $b_{\rm max}=50\,{\rm kpc}$ to characterize the cross-section in our estimate of Eq.~\ref{eq:phi_simplified}.

We compute Eq.~\ref{eq:phi_simplified} at $z=1.2$, a redshift at which several direct measurements of the \oii\ luminosity function are available (see below). We first use $C=1$, implying that all star forming galaxies are surrounded by \MgII\ gas. The resulting \OII\ luminosity function is shown in the left panel of Figure~\ref{plot_oii_mgii_connection} with the solid red line. For comparison, we show several estimates of the \OII\ luminosity function at the same redshift from narrow-band imaging surveys of \oiib\ emitting galaxies in the HST COSMOS 2 square degree field using the Suprime-Cam on the Subaru Telescope \citep{2007ApJS..172..456T} as well as from Ly et al. (2007). The overall agreement, both in shape and amplitude, is striking. The \MgII\ rest equivalent width distribution appears to be related to the \oii\ luminosity function, and therefore the p.d.f. of star formation rate. The connection between \MgII\ absorption and star formation is not only seen as a function of redshift, as shown through estimates of the \oii\ luminosity density $\mathcal{L}_{\rm OII}$, but appears to also operate as a function of $W_0$.

Comparing the red curve to the set of blue curves reveals a number of interesting properties: first of all we can see that the above relation maps \MgII\ absorbers with $1\lesssim W_0\lesssim6\,{\rm \AA}$ to \oii\ line luminosities ranging from $10^{41}$ to about $5\times10^{42}$ erg/s, i.e. the range corresponding to the exponential decline of the \oii\ luminosity function. Second, the match in amplitude shows that the \emph{incidence} of strong \MgII\ absorbers can be explained by the number density of \oii\ bright galaxies and finally the \emph{shape} of the two distributions are in good agreement, further indicating that these emission and absorption processes are related. These two quantities ultimately allow us to probe the same phenomenon: star formation.

In Eq.~\ref{eq:phi_simplified}, the terms ${\d^2 N}/{\d W_0\,\d z}$, $\alpha$ and ${\mathcal E}(z)$ are already constrained by direct observations and a set of cosmological parameters. Some constraints on the spatial extent of the cross-section $\sigma$ and the covering factor $C$ already exist but are less robust. Keeping in mind that the scatter in the reported values of $\Phi(L)$ only allows us to compare distributions within a factor $\sim2$, we now comment on these two quantities. Changing the value of the covering factor $C$ in Equations~\ref{eq:phi} and \ref{eq:phi_simplified} results in shifting the curves vertically. By definition, the amount of \oii\ luminosity density traced by \MgII\ absorbers must be lower than the total value, estimated from direct measurements. Depending on the chosen dataset, Figure~\ref{plot_oii_mgii_connection} shows that the amplitude of the \oii\ luminosity function implied by \MgII\ absorbers is either comparable to the measured $\Phi(L)$ or lower by about a factor two. This puts a strong constraint on the covering factor
\be
C=C_f\,C_\Omega\simeq0.5-1\,.
\ee
Therefore, the covering factor for \MgII\ absorption around star forming galaxies (with $L_{OII}\gtrsim10^{41}\,{\rm erg/s}$) is expected to be high, possibly close to unity. Using a different value of the cross-section $\sigma$ has several effects: it changes both the relation between $L$ and $W_0$, and the amplitude of the predicted $\Phi(L)$, shifting the prediction diagonally and changing its curvature. 
We find that only values of $20 \lesssim b_{\rm max}\lesssim 60\,{\rm kpc}$ do not over-predict the observed luminosity density at certain values of $L_{OII}$. The match between these two sets of curves is possible for only a restricted range of impact parameters, which encloses the mean value reported by direct observations of \MgII-galaxy associations.

We now look at the relation in Eq.~\ref{eq:phi} from the opposite direction, i.e. we use it as a model for the distribution of \MgII\ rest equivalent width. Using $C=1$, as discussed above, we can write
\be
\frac{\d^2 N}{\d W_0\,\d z} = \Phi(\log L)\;\frac{\alpha\,\sigma}{{\mathcal E}(z)}\,.
\ee
This relation provides a model explaining both the amplitude, shape and redshift-dependence of the \MgII\ rest equivalent width distribution, with no free parameter. The inferred distribution of \MgII\ rest equivalent widths is shown in the right panel of Figure~\ref{plot_oii_mgii_connection} with blue lines using the three measurements of the \oii\ luminosity function introduced above. For comparison, we show the measurement of ${\d^2 N}/{\d W_0\,\d z}$ from SDSS obtained by Nestor et al. (in prep.) for systems centered at $z=1.2$, with a half width of $\delta z=0.2$. This Figure allows us to illustrate the accuracy with which the latter distribution is measured with current datasets. The estimate of ${\d^2 N}/{\d W_0\,\d z}$ originates from the analysis of about 45,000 SDSS quasar spectra spread over a large fraction of the sky. It therefore does not suffer from cosmic variance.

While we have focused on the similarities between the shape and amplitude of $\Phi(\log L)$ and ${\d^2 N}/{\d W_0\,\d z}$, we now comment on possible departures between these quantities. As surveys become larger and selection effects better understood, a more robust estimate of the \oii\ luminosity function will allow us to explore possible deviations from the predictions given by Eq.~\ref{eq:phi}.
As discussed above, the ratio between the predicted and observed $\Phi(L)$ will provide us with an estimate of the covering factor for absorption around star forming galaxies. Deviations in shape could also indicate variations of the cross-section as a function of \oii\ luminosity and/or \MgII\ rest equivalent width. As mentioned above, the overall distribution of \MgII\ $W_0$ can be described by a Schechter function, with $W^\star\simeq0.6$ at $z=1$. Observational constraints exist down to about 0.1 \RAA. If the correspondence between $W_0$ and $L_{OII}$ holds in this regime, this would provide us with a mean of probing the \oii\ luminosity function down to values significantly fainter than current emission derived estimates at these redshifts.

In summary, we have shown that the distribution of \MgII\ rest equivalent widths and the \oii\ luminosity function track each other, as a function of both redshift and $W_0$, suggesting that these two observables probe the same phenomenon: star formation. This connection allows us to present the first non-parametric model for the distribution of \MgII\ rest equivalent widths, ${\d^2 N}/{\d W_0\,\d z}$, which appears to inherit its amplitude, shape and redshift dependence from the \OII\ luminosity function or equivalently the p.d.f. of star formation in the Universe.

\section{Discussion}\label{sec:discussion}

\subsection{Absorber-galaxy correlations}

As mentioned earlier, the nature of strong \MgII\ absorbers has been a matter of debate for more than two decades. One way to address this question has been to look for observational correlations between absorber and galaxy properties. However, until recently no compelling detection had been reported. This analysis has allowed us to show that \MgII\ absorbers trace a substantial fraction of the overall star formation rate of the Universe. This property actually sheds light on why previous attempts to detect correlations between $W_0({\rm MgII} )$ and galaxy secular properties have either failed, or reported rather weak trends (e.g. \citealt{K+2007}). As shown by Brinchmann et al. (2004) at $z<0.2$, the relation between SFR and galactic stellar mass (measured over five decades) is relatively shallow: $\rm SFR \propto M_*^{0.6}$. At high redshifts, observational constraints on the SFR-stellar mass relation are limited to massive galaxies (Noeske et al. 2007, and Daddi et al., 2007) but indicate a large scatter in SFR at a given mass. Therefore selecting galaxies with respect to their SFR, or equivalently $W_0$, will result in a sample with a wide range in stellar mass, broad band luminosity and colour. Correlations between absorber rest equivalent widths and galaxy secular parameters such as broad band luminosity or mass are thus expected to be weak, in agreements with results from \cite{Steidel+97,Zibetti+07,2009MNRAS.393..808M,2008ApJ...679.1218T} and \cite{2009ApJ...702...50G}.

\subsection{The nature of \MgII\ absorbers}
\label{sec:nature}

The nature of \MgII\ absorbers has been long been debated and so far no consensus has been reached regarding the physical process governing the observed equivalent widths. In this work we have shown the connection between the presence of \MgII\ absorbers in galactic halos and nearby star formation. Gas around a galaxy can be related to star formation in two different ways: it can be infalling and feeding an episode of star formation or it can be outflowing, due to star formation feedback processes such as supernova winds or radiation pressure.

Enhancements of star formation are expected to be triggered on a timescale of order the dynamical time ($\tau_d=R/v_c\sim10^8 $yr) following gas accretion. The time required for these systems to travel a distance $r$ of 20 kpc is
\be
\Delta t= 60\,{\rm Myr}\,\left( \frac{r}{20\,{\rm kpc}} \right) \,\Big( \frac{v}{300\,{\rm km\,s}^{-1}} \Big)^{-1}\,,
\label{eq:t}
\ee
where the value of the characteristic velocity is motivated by observations of blue-shifted \MgII\ absorption in the spectra of star-forming galaxies \citep{2009ApJ...692..187W} or using the observed \MgII\ rest equivalent width as a proxy for the projected velocity: $\Delta v \simeq 120\,(W_0/\RAA)\,$km/s (see section~\ref{sec:mgii}). Such velocities are however comparable to the circular velocity of $L^\star$ galaxies and can also be reached by infalling material. The above relation indicates that the timescale for the gas to travel is comparable to the timescale of a star formation episode. Note that we would not expect any correlation between $W_0$ and instantaneous star formation rate if $\tau_d\ll\Delta t$. 

The high covering factor around star forming galaxies implied in our analysis allows us to shed light on the direction of motion of the gas. Although there is currently no direct observational evidence for infalling cold gas onto galaxies, results from numerical hydrodynamic simulations suggest that gas accretion leading to efficient star formation mostly occurs through cold flows, i.e. dense filamentary structures penetrating through the diffuse gaseous halos of galaxies \citep{Keres05,2009Natur.457..451D}. Such models suggest that these elongated structures give rise to covering factors of order 10-20\%. In Section~\ref{sec:origin} we showed that, in order not to over-predict the observed \oii\ luminosity function, the covering factor for \MgII\ absorption around star forming galaxies is necessarily high, with $C=C_f\,C_\Omega\gtrsim0.5$. While such a value appears somehow too high for cold flows, it is naturally reproduced by outflows, which are known to take place with a large opening angle \citep{2005ARAA..43..769V}. 

Another argument in favor of outflows comes from the similarities between the results presented in this study, i.e. the connection between intervening \MgII\ absorbers and star formation, and the recent results by \cite{2009ApJ...692..187W} who showed that blue-shifted \MgII\ absorption, i.e. outflowing gas, is ubiquitous in star-forming galaxies at $z\sim1$. These authors found the presence of outflows in all subdivisions of their sample of star-forming galaxies, covering a factor ten in SFR and $B$-band luminosity, and a factor 30 in stellar mass. 
Their blue-shifted \MgII\ systems appear to share all the properties of the intervening systems used in our study: rest equivalent widths, redshift range, high covering factor and especially the correlation with star formation.
Occam's razor (c. 1320), which tells us that entities should not be multiplied unnecessary, leads us to postulate that the nature of the \MgII\ gas is the same in both types of analyses. In other words, blue-shifted \MgII\ absorbers seen in self absorption and intervening \MgII\ absorbers might be the same gas clouds. This implies that the blue-shifted absorption seen in star forming galaxies (e.g. \citealt{1990ApJS...74..833H,1999ApJ...513..156M, 2009ApJ...692..187W} may well extend to radii up to tens of kpc.

\subsection{A new probe of star formation}

Our analysis has shown that the bulk of star formation is traced by \MgII\ absorbers. This connection provides us with a new tool to explore star formation in the Universe, over cosmological times. Here we discuss its potential and highlight differences with existing techniques.
\begin{itemize}
\item Absorption measurements offer a number of advantages: the detectability of absorption lines does not depend on redshift (while galaxy fluxes drop rapidly at high redshifts) and perhaps more importantly, the observed properties of absorption lines are not sensitive to dust extinction. \MgII\ absorbers can therefore provide us with a powerful and complementary tracer of star formation. 
\item Absorber systems allow us to apply emission line studies in a noise-dominated regime. Only a handful of low-redshift \MgII\ absorbers allow for the direct detection of \oii\ emission lines in individual systems. For the vast majority, no direct detection is possible in SDSS quasar spectra. However, being able to average over thousands of lines-of-sights allows us to estimate the mean \oii\ luminosity of the corresponding population, down to emission levels comparable to what is currently obtained with some of the deepest narrow band surveys.
\item This technique does not suffer from contamination from misidentified emission lines, a severe limitation for narrow-band imaging surveys of \oiib\ emission (see \citealt{2007ApJS..172..456T,2007ApJ...657..738L}).
\item In optical spectra \MgII\ absorbers are detected from the ground over the range $0.3<z<2.2$. Infrared spectroscopy can extend this redshift path up to the reionization epoch. For example, the upcoming near-infrared spectrograph FIRE \citep{2008SPIE.7014E..27S} will allow the detection of \MgII\ absorption and \oii\ emission up to $z\sim6$, for which the number of quasars will become the limiting factor. Infrared spectroscopy will allow us to map the star formation history beyond redshift two, \emph{with the same tracer}, and provide a new set of constraints on the decline of the SFR density at high redshift.
\end{itemize}

A vast amount of \MgII\ data has been accumulated over the years. Its analysis has mainly focused on attempting to reveal the nature of these systems. We can now use these datasets to obtain new insights into star formation, over cosmological times. About 16,000 \MgII\ absorbers have been detected in SDSS quasar spectra \cite{Quider}. The analysis of the entire sample (DR7) will double this number. The details of \MgII\ absorbers (number of components, velocity widths, etc.) accessible through high-resolution spectroscopy can be used to put constraints on star formation processes. 
Understanding this type of information might allow us to constrain the recent star formation history of galaxies. Such a technique will be complementary to spectral energy distribution analyses.



More work needs to be done in order to understand the origin of the empirical \oii-\mgii\ scaling relation and especially its redshift dependence.  If the redshift dependence of $\sloii(W_0)$ can be explained, \MgII\ absorbers can then, alone, be used to trace star formation in the Universe. Finally, comparing SFR$(z)$ from emission and absorption techniques might shed light on the redshift dependence of the dust extinction affecting emission based estimators. With appropriate wavelength coverages, similar absorption-emission scaling relations can be explore, for example between \MgII\ absorption and H-$\alpha$ emission. It will be interesting to extend the current analysis to other absorption transitions  (FeII, CIV, etc.) and test their connection to the process of star formation.

\section{Summary}

We present an empirical connection between cold gas in galactic halos and star formation. Using a sample of more than 8,500 \MgII\ absorbers from SDSS quasar spectra, we report the detection of a $15\,\sigma$ correlation between the rest equivalent width $W_0$ of \MgII\ absorbers and the associated \oii\ luminosity, an estimator of star formation rate. 
This correlation allows us to show that
\begin{enumerate}
\item \MgII\ absorbers trace a substantial fraction of the \oii\ luminosity density in the Universe, i.e. global star formation. The SFR density probed by \MgII\ absorbers recovers the overall star formation history of the Universe derived from classical emission estimators up to $z\sim2$. 
\item the distribution function of \MgII\ rest equivalent widths, $\d N/\d W_0$  inherits both its shape and amplitude from the \oii\ luminosity function $\Phi(L)$. These distributions can be naturally connected, without any free parameter. 
\item the covering factor for cold gas around star forming galaxies appears to be high: $C\gtrsim0.5$. The determination of this value is currently limited by the scatter in the reported estimates of \oiib\ luminosity densities.
\end{enumerate}
These results not only shed light on the nature of \MgII\ absorbers but also provide us with a new probe of star formation, in absorption, i.e. in a way which does not suffer from dust extinction and with a redshift-independent sensitivity. As shown in this analysis, such a tool can be applied in a noise-dominated regime, i.e. using a dataset for which emission lines are not detected in individual objects. This is of particular interest for high redshift studies.

We argue that outflows appear as a favored explanation regarding the nature of \MgII\ absorbers and that 
blue-shifted \MgII\ absorption seen in the spectra of star forming galaxies could be the same systems, implying that the observed outflowing gas can reach radii of $\sim50$ kpc.

\section*{Acknowledgements}

We thank David Turnshek and Sandhya Rao for their crucial role the compilation of the \MgII\ absorber catalog based on SDSS data. We are grateful to the authors of the \cite{2007ApJ...657..738L} for having provided us with the values of their observed \oii\ luminosity function. We thank Doron Chelouche, Jason Prochaska and Norman Murray for useful discussions.

Funding for the Sloan Digital Sky Survey (SDSS) has been provided by the Alfred P. Sloan Foundation, the Participating Institutions, the National Aeronautics and Space Administration, the National Science Foundation, the U.S. Department of Energy, the Japanese Monbukagakusho, and the Max Planck Society. The SDSS Web site is http://www.sdss.org/.

The SDSS is managed by the Astrophysical Research Consortium (ARC) for the Participating Institutions. The Participating Institutions are The University of Chicago, Fermilab, the Institute for Advanced Study, the Japan Participation Group, The Johns Hopkins University, the Korean Scientist Group, Los Alamos National Laboratory, the Max-Planck-Institute for Astronomy (MPIA), the Max-Planck-Institute for Astrophysics (MPA), New Mexico State University, University of Pittsburgh, University of Portsmouth, Princeton University, the United States Naval Observatory, and the University of Washington.

\section*{Appendix: fiber aperture correction}
\label{appendix_fiber}

\begin{figure}
\begin{center}
  \includegraphics[width=.8\hsize]{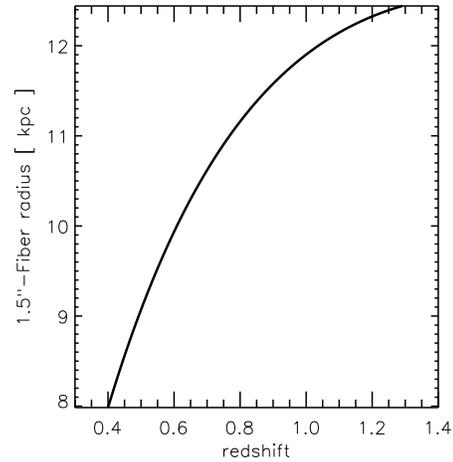}
  \caption{Physical radius mapped by the SDSS spectroscopic fiber, as a function of redshift}
\label{plot_loii_fiber}
\end{center}
\end{figure}

\end{document}